\documentstyle[prl,aps]{revtex}

\begin{document}
\draft

\twocolumn[\hsize\textwidth\columnwidth\hsize
\csname@twocolumnfalse\endcsname

\title{Incommensurate magnetism in cuprate materials}
\author{F. Mancini and D. Villani}
\address{Dipartimento di Scienze Fisiche ''E. R. Caianiello'' - Unit\`a I.N.F.M. di
Salerno\\
Universit\`a di Salerno, 84081 Baronissi (SA), Italy}
\author{H. Matsumoto}
\address{Institute for Material Research, Tohoku University, Sendai 980, Japan}
\date{December 16, 1996}
\maketitle

\begin{abstract}\widetext
In the low doping region an incommensurate magnetic phase is observed in $%
La_{2-x} Sr_x CuO_4$. By means of the composite operator method we show that
the single-band 2D Hubbard model describes the experimental situation. In
the higher doping region, where experiments are not available, the
incommensurability is depressed owing to the van Hove singularity near the
Fermi level. A proportionality between the incommensurability amplitude and
the critical temperature is predicted, suggesting a close relation between
superconductivity and incommensurate magnetism.
\end{abstract}

]

\narrowtext

The dynamical spin susceptibility for cuprate materials has been
investigated by inelastic neutron scattering and NMR techniques. Neutron
scattering data on $La_{2-x}(Ba,Sr)_xCuO_4$ have shown [1-7] that away from
half-filling the commensurate antiferromagnetic phase is suppressed and
short-range incommensurate antiferromagnetism develops. The magnetic Bragg
peak in the dynamical structure factor $S({\bf k},\omega )$ broadens and
develops a structure with four peaks located at $[(1\pm \delta )\pi ,\pi ]$
and $[\pi ,(1\pm \delta )\pi ]$. The incommensurability amplitude $\delta (x)
$ does not depend on the frequency and temperature; it is zero up to the
doping $x\approx 0.05$ where a commensurate-incommensurate transition is
observed, then, increases with the hole concentration $x$, following the
linear law $\delta (x)\approx 2x$ up to $x\approx 0.12$; beyond this point
there is a deviation downwards [7]. Unfortunately, experimental data above $%
x=0.18$ are not available, due to the difficulty in preparing single
crystals. It is important to stress [5] that the value of doping $x=0.05$,
where the transition is observed, corresponds to the value of doping where
the material becomes superconducting. These incommensurate spin fluctuations
are not observed in other cuprate materials; a flat topped magnetic peak has
been observed [8] in $YBa_2Cu_3O_{6+y}$ with $y\approx 0.6$, while for the
case of the electron-doped $Nd_{2-x}Ce_xCuO_4$ no incommensurate magnetism
has been observed [9]. The difference in the spatial modulation
experimentally observed in $La_{2-x}Sr_xCuO_4$ and $YBa_2Cu_3O_{6+y}$ has
been related to a difference in the topology of the Fermi surface [10,11].

From a theoretical side the presence of incommensurate magnetic correlations
was firstly found in the 2D Hubbard [12] and in the $t-J$ model [13] by
Quantum Monte Carlo (QMC) calculations. To improve the situation the $%
t-t^{\prime}$ Hubbard model has been considered, but the results are not
definite and there is no general agreement [14].

In this Letter we shall study the spin magnetic susceptibility of the
two-dimensional single-band Hubbard model by means of the Composite Operator
Method (COM) [15-16]. We have shown [16] that in the static approximation
the dynamical spin susceptibility is given by the following expression: 
\begin{eqnarray}
\chi ({\bf k},\omega ) &=&{\frac 2{n^2-n-2D}}\left[ n\left(
Q_{1111}^R+2Q_{1112}^R+Q_{1212}^R\right) \right.   \nonumber \\
&&\left. +(2-n)\left( Q_{1212}^R+2Q_{1222}^R+Q_{2222}^R\right) \right] 
\end{eqnarray}
where $n$ is the particle density, D is the double occupancy and by $%
Q_{\alpha \beta \gamma \delta }^R({\bf k},\omega )$ we mean the retarded
part of $Q_{\alpha \beta \gamma \delta }({\bf k},\omega )=i\int {\frac{%
d^2pd\Omega }{(2\pi )^3}}G_{\alpha \beta }({\bf k}+{\bf p},\omega +\Omega
)G_{\gamma \delta }({\bf p},\Omega )$. The 2x2 matrix $G({\bf k},\omega )$
is the thermal causal Green's function, defined by $G({\bf k},\omega
)=\langle T[\psi (i)\psi ^{\dagger }(j)]\rangle _{F.T.}$, where $\psi (i)$
is the doublet composite operator 
\begin{equation}
\psi (i)={%
{\xi (i) \choose \eta (i)}
}
\end{equation}
with $\xi _\sigma (i)=c_\sigma (i)\left[ 1-n_{-\sigma }(i)\right] $ and $%
\eta _\sigma (i)=c_\sigma (i)n_{-\sigma }(i)$. $c_\sigma (i)$ is the
electron operator at site $i$. By means of the equation of motion and by
considering the static approximation, where finite lifetime effects are
neglected, the Green's function $G({\bf k},\omega )$ can be computed in the
course of a fully self-consistent calculation where no adjustable parameters
are considered [15].

The calculation of the uniform static susceptibility $\chi_0(x,T)$ for
various values of doping and temperature has been given in Ref. 16, where we
showed that Eq. (1) qualitatively reproduces the experimental situation
observed in $La_{2-x} Sr_x Cu O_4$. The principal results obtained in Ref.
16 can be so summarized. For a fixed temperature, $\chi_0(x,T)$ is an
increasing function of the doping, reaches a maximum at a critical doping $%
x_c$, then decreases. The value of $x_c$ does not change with temperature
and is determined by the ratio $U/t$, varying from 0 to 1/3 when $U/t$
changes from zero to infinite; for $U/t=4$ we have $x_c=0.27$. This doping
dependence qualitatively reproduces the experimental behavior observed in $%
La_{2-x} Sr_x Cu O_4$ [17], where a critical value $x_c\approx 0.25$ is
reported. For a given doping, when the system is away from the critical
density, $\chi_0(x,T)$ as a function of temperature has a behavior similar
to a 2D Heisenberg antiferromagnet with a maximum at a certain temperature $%
T_m$; the position of $T_m$ decreases as the system is doped away from
half-filling and tends to zero for $x\to x_c$; in the vicinity of the
critical doping there is a large increase of $\chi_0(x,T)$ for low
temperatures. The behaviors of $T_m$ as a function of $x$ and of $\chi_0(x,T)
$ as a function of $T_m$ well reproduce the experimental data of Ref. 17.

The peak exhibited by $\chi_0(x,T)$ for a certain critical doping is related
to the fact that for $x=x_c$ the Fermi energy crosses the vHs. This can be
seen in Fig. 1, where $N(E_F)$, the density of states calculated at the
Fermi energy, and $\chi_0(x,T)$, the uniform static susceptibility, are
given versus the doping parameter $x$. We have chosen $U/t=4$ and $k_B
T/t=0.01$. At half-filling the Fermi energy is at the center of the two
Hubbard bands; by varying the dopant concentration some weight is
transferred from the upper to the lower band, $E_F$ moves to lower energies
and crosses the vHs for a critical value of the doping; further increasing $x
$, $E_F$ moves away from the vHs. A study of the Fermi surface shows that
for $x>x_c$ we have a closed surface which becomes nested at $x=x_c$ and
opens for $x<x_c$. An enlarged Fermi surface with a volume larger than the
noninteracting one has been reported by QMC calculations [18,19] and by
other theoretical works [20].

To understand the role played by the vHs in the case of spin fluctuations,
it is useful at first to consider the case of noninteracting Hubbard model
(i.e. $U=0$). What we learn [21] from the study of this model can be so
summarized.

\noindent - The ${\bf k}$-dependent susceptibility $\chi ({\bf k})$ exhibits
a maximum at a certain value ${\bf k}^{*}$ which depends on the position of
the vH energy $\omega _{vH}$ with respect to the Fermi energy $E_F$.

\noindent - When $n=1$ $\omega _{vH}=E_F$ and ${\bf k}^{*}={\bf Q}=(\pi ,\pi
)$. In addition there is a singularity coming from the nesting of the Fermi
surface and the staggered susceptibility $\chi ({\bf Q})$ exhibits a
stronger divergence than $\chi _0$.

\noindent - When $n\neq 1$ the Bragg peak at ${\bf Q}=(\pi ,\pi )$ opens in
four peaks, situated at ${\bf k}^{*}=[\pi (1\pm \delta ),\pi (1\pm \delta )]$%
; there is a transition from commensurate to incommensurate magnetism. The
incommensurability amplitude $\delta (x)$ increases as a function of the
doping $x$ with the same law as the shifting of $E_F$ with respect to the
vHs: 
\begin{equation}
\begin{array}{ll}
\omega _{\nu H}-E_F\approx ax^{4/3}\qquad  & a\approx 3.62 \\ 
\delta (x)\approx bx^{4/3}\qquad  & b\approx 1.31
\end{array}
\end{equation}

In the interacting case, the shifting of the vHs and the band structure have
a drastic influence on the form of the susceptibility. Calculations based on
the use of Eq. (1) show that around ${\bf Q}=(\pi ,\pi )$ $\chi ({\bf k})$
has an incommensurate structure along the four corners of a square, with a
minimum at ${\bf Q}$ . This incommensurate structure contains a mixing of
two components. The relative position and the intensity of the two
contributions change significantly with doping. To study the effect of the
interaction we see that the $k$-dependent susceptibility $\chi ({\bf k})$
can be written as $\chi ({\bf k})=\sum_{i,j=1}^2\chi _{ij}({\bf k})$ where 
\begin{equation}
\chi _{ij}({\bf k})=\int {\frac{d^2p}{(2\pi )^2}}{\frac{f[E_i({\bf k}+{\bf p}%
)]-f[E_j({\bf p})]}{E_i({\bf k}+{\bf p})-E_j({\bf p})}}K_{ij}({\bf k},{\bf p}%
) \\
\end{equation}
The quantities $K_{ij}({\bf k},{\bf p})$ are expressed in terms of the
spectral intensities. The term $\chi _{inter}=\chi _{12}+\chi _{21}$
describes transitions between the two bands $E_1({\bf k})$, the upper
Hubbard band, and $E_2({\bf k})$, the lower Hubbard band; while the two
terms $\chi _{11}$ and $\chi _{22}$ describe intraband transitions. Since $%
E_1$ takes values mostly above the chemical potential, the contribution of $%
\chi _{11}$ is small. The interband term $\chi _{inter}$ is reported in Fig.
2. This term originates a peak in the susceptibility, which moves from the
commensurate position ${\bf Q}=(\pi ,\pi )$ to $(\pi ,\pi /2)$ when the
doping is increased from $x=0$ to $x=x_c$. The intensity of the peak
decreases by increasing doping. The intraband term $\chi _{22}$ is reported
in Fig. 3. This term gives a peak which is a reminiscent of the Van Hove
singularity in the density of states. At zero doping the vHs is far from the
Fermi energy and the peak is located $(\pi /\pi /2)$ and has a low
intensity. When doping increases, the peak increases its intensity and moves
along the line ($k_x=\pi ,\pi /2\le k_y<3\pi /2$). At the critical doping $%
x=x_c$ the vHs lies on the Fermi energy and the Fermi surface is nested.
Then, the peak of $\chi _{22}$ is situated at ${\bf Q}$ and has a very high
intensity, due to the concomitance of these two effects. It is interesting
to note that the peak position of $\chi _{22}$ moves towards ${\bf Q}$ with
the same law as given in Eq. 3.

The total susceptibility is reported in Fig. 4 for three values of doping.
For zero doping we mainly have a commensurate structure with a peak coming
from $\chi _{inter}$, located at $(\pi ,\pi )$, and a smaller peak, coming
from $\chi _{22}$, located near $(\pi ,\pi /2)$. Upon doping, the two peaks
moves for different reasons. $\chi _{22}$ moves because the Van Hove
singularity moves towards the Fermi energy. $\chi _{inter}$ moves because
the band structure changes with doping. When the critical doping $x=x_c$ is
reached, the Van Hove singularity is at the Fermi energy and the Fermi
surface is nested. A commensurate structure is recovered with a very high
peak coming from $\chi _{22}$.

In Fig. 5 the incommensurability amplitude $\delta (x)$ is reported as a
function of doping. In the region of low (high) doping the peak coming from $%
\chi _{inter}$ ($\chi _{22}$) is predominant and very well separated from
the other; in these regions $\delta (x)$ has been evaluated as the middle
point of the half-width of the peak. In the region $0.10\le x\le 0.18$ the
two peaks overlap and $\delta (x)$ has been calculated by taking the average
of both peaks and we have a plateau due to the superimposition of $\chi _{22}
$ and $\chi _{inter}$. For comparison we report the experimental data of
Refs. 4, 5 and 7. The linear behavior of $\delta (x)$, observed in the low
doping region, agrees exceptionally well with the experimental data,
reported in Refs. 4-7; the downward deviation reported in Ref. 7 for $x>0.12$
might correspond to the plateau theoretically observed. One of the most
striking feature of the results presented in Fig. 5 is the similarity
between the incommensurability amplitude $\delta (x)$ and the critical
temperature $T_c$. $\delta (x)$ is maximum in the region of optimal doping
where $T_c$ is maximum. It has already been observed in Ref. 7 that there is
a linear relation between $\delta (x)$ and $T_c$ up to the optimal doping
level $x\simeq 0.15$. Our theoretical results extend to the all region of
doping a relation of proportionality between $\delta (x)$ and $T_c$.

The same result for $\delta (x)$ can be obtained by considering Im $\chi (%
{\bf k},\omega )$. Some results have been given in Ref. 22. We preferred to
study the ${\bf k}$-dependent susceptibility $\chi ({\bf k})$ because this
quantity provides more strict information about the spatial range of the
magnetic correlations. On the other hand an exact experimental determination
of $\chi ({\bf k})$ is not easy, since it must be calculated by the
accessible $S({\bf k},\omega )$ through a Kramers-Kronig relation.

The present analysis shows that the interaction in the Hubbard model has
mainly two effects. One is the change of the critical doping from $x=0$ to
some critical $x_c$, due to the shift of the vHs. This shift explains and
well reproduces the unusual normal state behavior of $\chi_0$ in hole-doped
cuprates. The other is a band structure effect which is responsible of the
incommensurate modulation of the magnetic susceptibility in the low doping
region.

The picture that emerges is that the magnetism probed by neutron scattering
experiments is correlated with the carrier density. In the low doping region
the susceptibility is mainly controlled by the term $\chi _{inter}$ which
describes band structure effects and then reflects the topology of the Fermi
surface. In the overdoped region the Fermi energy is close to the vHs and
the effect of nesting in the intraband term is important. In $%
YBa_2Cu_3O_{6+y}$ we have a different topology of the Fermi surface and no
nesting is expected; this might be the reason why incommensurability is not
observed.

The main results obtained in this Letter can be so summarized. There is
experimental evidence that in hole-doped high $T_c$ cuprates the Fermi level
is close to the vHs for values of doping close to those where the
superconducting phase is suppressed. In the context of the Hubbard model a
van Hove scenario well describes some of the unusual properties observed in
the normal state, but an analysis show that this scenario is related to the
overdoped region and not to the optimal doping. The existence of a critical
doping where the vHs lies on the Fermi energy should imply a peak in the
staggered susceptibility. Then, we predict that commensurate magnetism
should be recovered in the nearness of the critical doping, implying a
proportionality relation between the incommensurability amplitude $\delta(x)$
and the superconducting critical temperature. Recalling that in $La_{2-x}
Sr_x Cu O_4$ the commensurate-incommensurate transition is observed at the
same value of doping $x\simeq 0.05$ where superconductivity starts, at least
for $La_{2-x} Sr_x Cu O_4$, a scenario [23] which relates the
superconducting phase to the presence of incommensurate magnetism emerges.

\acknowledgments
The authors wish to thank Prof. M. Marinaro and Dr. A. Avella for many
valuable discussions.

\centerline{\bf Figure Captions}

Fig. 1 The density of states at the Fermi energy $N(E_F)$ and the uniform
static susceptibility $\chi_0(x,t)$ as functions of the doping $x$. $U/t=4$
and $k_B T/t=0.01$.

Fig. 2 The interband term $\chi _3({\bf k})$ along the line ${\bf k}=(\pi
,k_y)$ for $k_BT/t=0.01$ and for various values of the doping $x\le 0.27$
with step $0.03$. $U/t=4$.

Fig. 3 The intraband term $\chi _2({\bf k})$ along the line ${\bf k}=(\pi
,k_y)$ for $k_BT/t=0.01$ and for various values of the doping $x\le 0.27$
with step $0.03$. $U/t=4$.

Fig. 4 The spin magnetic susceptibility $\chi ({\bf k})$ along the line $%
{\bf k}=(\pi ,k_y)$ for various values of the doping $x$. $U/t=4$ and $%
k_BT/t=0.01$ .

Fig. 5 The incommensurability amplitude $\delta(x)$ as a function of the
doping $x$. The dashed line indicates the theoretical result. $U/t=4$ and $%
k_BT/t=0.01$.

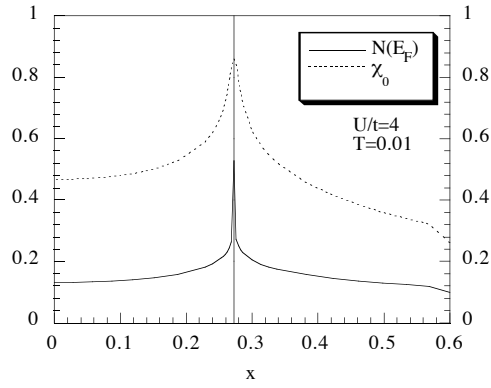
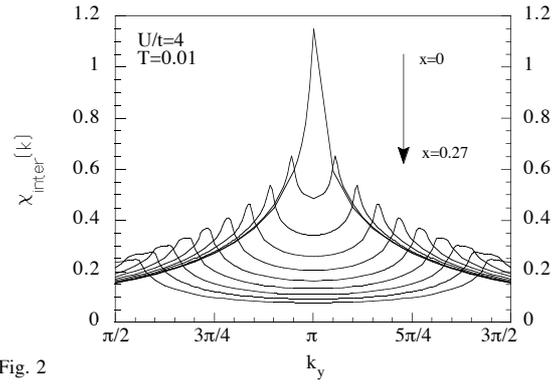
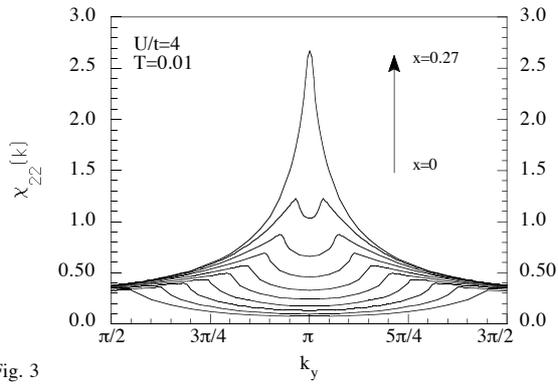
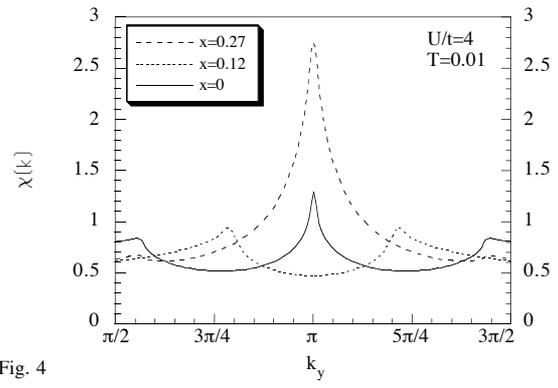
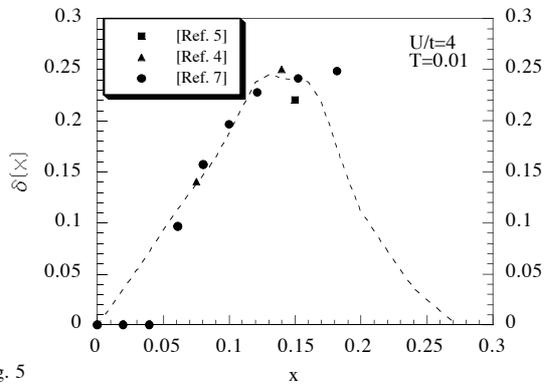
\end{document}